\documentclass[journal, 10pt]{IEEEtran}

\ifCLASSINFOpdf
\else
\fi
\usepackage{cite}
\usepackage{graphicx}
\usepackage{float}
\usepackage[small]{caption}
\usepackage{subcaption}
\usepackage{color,soul}
\usepackage{array}
\usepackage{amsmath}
\usepackage{booktabs}
\usepackage{footnote}
\makesavenoteenv{tabular}
\makesavenoteenv{table}
\usepackage{color}
\usepackage{amssymb}
\usepackage[linesnumbered,ruled,vlined]{algorithm2e}

\usepackage{algpseudocode}

\usepackage{setspace}

\newcommand\tab[1][1.5cm]{\hspace*{#1}}
\begin{document}
\pagenumbering{gobble}
\title{Approximate Logic Synthesis: A Reinforcement Learning-Based Technology Mapping Approach}

\author{ Ghasem Pasandi, Shahin Nazarian, and Massoud Pedram \\
  Department of Electrical and Computer Engineering\\ University of Southern California, Los Angeles, CA 90089.\\\{pasandi, shahin.nazarian, pedram\}@usc.edu

%\thanks{Manuscript submitted November 20, 2015.}
}


\maketitle

\begin{abstract} 
Approximate Logic Synthesis (ALS) is the process of synthesizing and mapping a given Boolean network to a library of logic cells so that the magnitude/rate of error between outputs of the approximate and initial (exact) Boolean netlists is bounded from above by a predetermined total error threshold. In this paper, we present Q-ALS, a novel framework for ALS with focus on the technology mapping phase. Q-ALS incorporates reinforcement learning and utilizes Boolean difference calculus to estimate the maximum error rate that each node of the given network can tolerate such that the total error rate at non of the outputs of the mapped netlist exceeds a predetermined maximum error rate, and the worst case delay and the total area are minimized. Maximum Hamming Distance (MHD) between exact and approximate truth tables of cuts of each node is used as the error metric. In Q-ALS, a Q-Learning agent is trained with a sufficient number of iterations aiming to select the fittest values of MHD for each node, and in a cut-based technology mapping approach, the best supergates (in terms of delay and area, bounded further by the fittest MHD) are selected towards implementing each node. Experimental results show that having set the required accuracy of 95\% at the primary outputs, Q-ALS reduces the total cost in terms of area and delay by up to 70\% and 36\%, respectively, and also reduces the run-time by 2.21$\times$ on average, when compared to the best state-of-the-art academic ALS tools. 
\end{abstract}
%%%%%%%%%%%%%%%%%%%%%%%%%%%%%%%%%

\IEEEpeerreviewmaketitle

\section{Introduction}
\label{Intro:sec}
\setstretch{0.98}
Approximate computing is defined as a computing technique to generate results with possible inaccuracy. This happens by relaxing the exact equivalency requirements between provided specifications and generated results. Approximate computing  has attracted tremendous attentions in many different fields that can tolerate inaccuracy, such as video and image processing \cite{huang2015surface}, search engines \cite{ranjan2016approximate,yang2003approximate}, machine learning \cite{smola2000sparse,gal2016dropout,rezende2014stochastic, li2017normalization, pasandi2018TruthNet}, and approximate hardware design \cite{dutt2016comparative,masadeh2018comparative, mrazek2016design} by providing improvement in speed and saving in resources. For example,  Esmaeilzadeh et. al \cite{Esmaeilzadeh:2012_1} accelerated neural networks using approximate computing techniques by introducing a parrot transformation and a neural processing unit. 

Approximate computing can be done in different abstraction levels and transformations, one of which is the logic synthesis. Logic synthesis is the process which optimizes and maps a given Boolean network into a netlist consisting of logic gates. The process of exact mapping is present in a variety of synthesis tools. However, in recent times as the demand for energy and area deficiencies has been increasing, there is a need for an alternative approach to meet this demand. This can be achieved by identifying the digital systems which can tolerate errors and arriving at a mapping process which sacrifices accuracy to achieve improvements in area, energy, or delay. This process of synthesis is called Approximate Logic Synthesis (ALS). 

There are several papers in the literature for ALS including SALSA \cite{venkataramani2012salsa}, SASIMI \cite{venkataramani2013substitute}, selection based (Single-Selection and Multi-Selection) algorithms \cite{wu2016efficient} and many more (see Section \ref{prior:sec}), which present new algorithms and/or synthesis tools for generating approximate netlists bounded by a predetermined error rate or error magnitude. These approaches typically suffer from large run-time values. Furthermore, they lack powerful machine learning engines to learn from previous optimization process steps, to better optimize for power, area or delay of the circuit. In this paper, we present Q-ALS, a novel approach for ALS which is based on Reinforcement Learning (RL) algorithms. 

Q-ALS embeds a Q-learning agent, which is trained and utilized for determining the maximum error that can be tolerated by a node in a given network such that the total error rate at non of the primary outputs of the network exceeds a predetermined maximum error rate. The Hamming distance between truth tables of exact and approximate implementations of a node is used as the error metric. To estimate the total error rate at primary outputs which is resulted from approximation in a node, a probabilistic approach based on Boolean difference calculus is utilized. Experimental results verify that Q-ALS provides considerable QoR (Quality of Results) improvements in terms of run-time, total area, and the worst case delay of many tested benchmark circuits over other ALS frameworks. Due to the high impact of technology mapping on the final delay and area results, the focus of approximation in this paper is on the technology mapping phase. To the best of our knowledge, this paper is the first to address the problem of approximate logic synthesis using RL algorithms. 

The rest of the paper is organized as follows: Section \ref{prior:sec} summarizes some of the previous works on ALS. Section \ref{back:sec} provides a quick background on Q-learning and cut-based technology mapping, which will be used in the rest of the paper. Section \ref{prop:sec} presents our Q-ALS framework. Experimental results on many benchmark circuits and conclusions appear in Sections \ref{exp:sec} and \ref{conc:sec}, respectively.
%%%%%%%%%%%%%%%%%%%%%%%%%%%%%%%%%%%%%%%%%%%%%%%%%%%%%%%%%%%%%%%%%%%%%%%%%%%%%%%%%%%%%%%%%%%%%%%%%%%%%%%%%%%%
\section{Related Work}
\label{prior:sec}
Miao \textit{et al.} \cite{miao2013approximate} formulated the ALS problem unconstrained by error rate as a Boolean relations (BR) minimization problem, further refined by a heuristic approach to satisfy the error frequency constraints. Shin and Gupta \cite{shin2010approximate} developed a method to reduce the literal count by exploiting the error tolerance rate during the circuit design. Miao \textit{et al.} \cite{miao2014multi} presented a method to reduce the gate count of a given circuit by using the external don't care (EXDC) sets that maximally approach the Boolean relation with compliance to the constrained error magnitude. A novel ALS framework was presented in \cite{liu2017statistically} that performs statistical testing to certify the area optimized circuits with high confidence under the given error constraint. This is done by continuously monitoring the quality of the generated design. In \cite{venkataramani2012salsa}, the existing synthesis tools such as SIS \cite{sentovich1992sis} and Synopsys Design Compiler \cite{compiler2001synopsys} are utilized to perform area reduction by using approximate don't cares (ADCs) adhering to the given quality bounds. An ALS tool called SASIMI is presented in \cite{venkataramani2013substitute}, which works with multiple error metrics such as error rate and error magnitude. SASIMI aims for area and power reduction by substituting the signal pairs which accept the same values with high probability. Wu and Qian \cite{wu2016efficient} addressed the ALS problem for multi-level circuits by reducing the size of nodes in the given Boolean network with the help of approximating their factored-forms. They presented two algorithms namely Single-Selection and Multi-Selection with similar area savings but better run-time for the latter one.

There have been a few recent publications on the utilization of machine learning algorithms in Electronic Design Automation (EDA). In \cite{haaswijk2018deep}, deep reinforcement learning advances are employed in logic optimization. More specifically, the logic optimization problem is formulated as a deterministic Markov decision process, and a generic algorithm is presented to solve it. In \cite{beerel2018opportunities}, several classical Computer-Aided Design (CAD) algorithms are presented, which can benefit from advances in machine learning. The process of finding the error constraint of each node for large networks leads to diminished performance. The framework which we present in this paper tackles the performance degradation problem by utilizing RL algorithms to find the maximum tolerable error rate at each node. RL, which is shown to resolve such issues \cite{ipek2008self}, will lead to a superior performance in terms of speed, area, and accuracy.
%%%%%%%%%%%%%%%%%%%%%%%%%%%%%%%%%%%%%%%%%%%%%%%%%%%%%%%%%%%%%%%%%%%%%%%%%%%%%%%%%%%%%%%%%%%%%%%%%%%%%%%%%%%%%%%
%%%%%%%%%%%%%%%%%%%%%%%%%%%%%%%%%%%%%%%%%%%%%%%%%%%%%%%%%%%%%%%%%%%%%%%%%%%%%%%%%%%%%%%%%%%%%%%%%%%%%%%%%%%%%%%%%%%%%%%
\begin{figure*}[t]
        \centering
        \begin{subfigure}[!t]{0.56\textwidth}
                \centering
                \includegraphics[width=\textwidth]{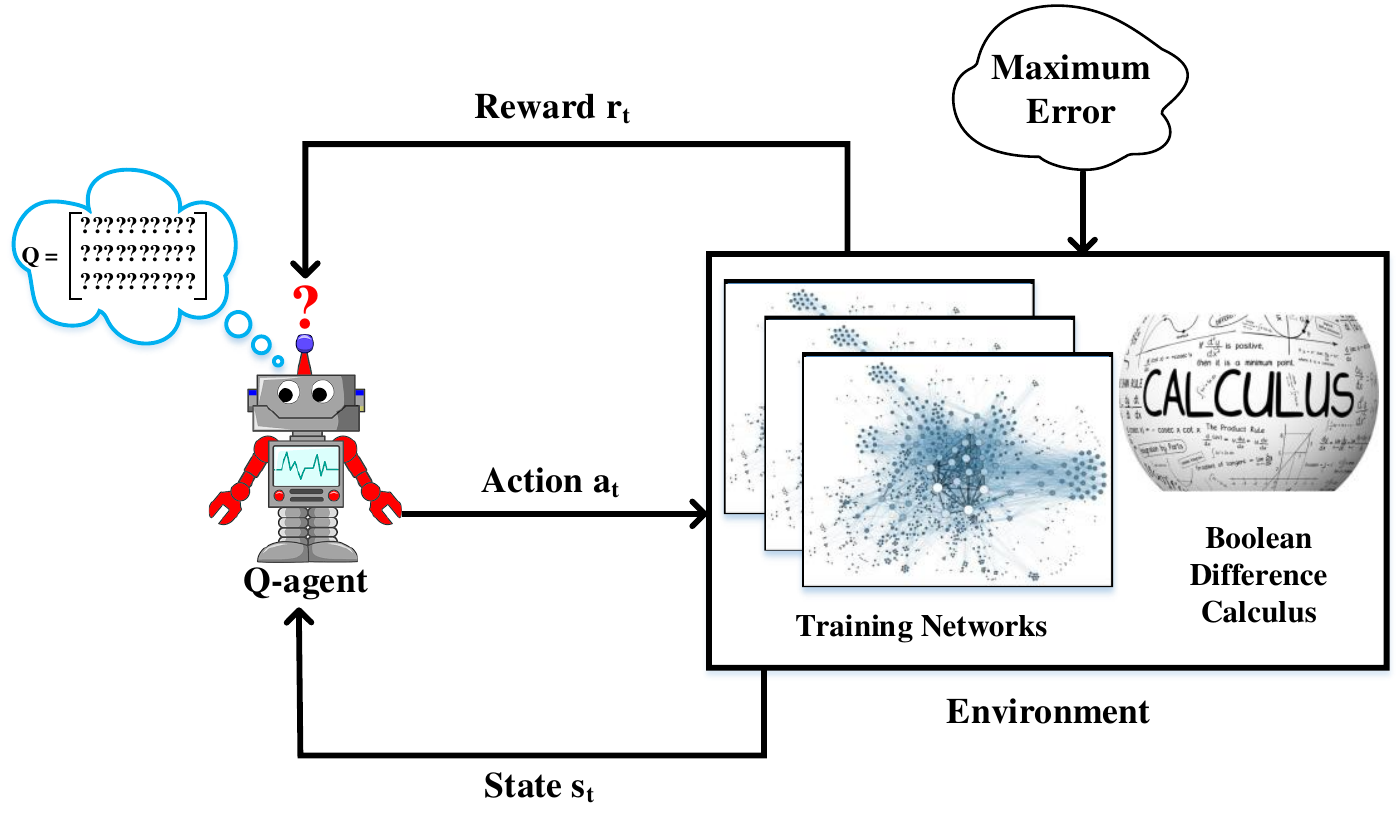}
                \caption{}
                \label{flow_a}
        \end{subfigure}
        %%%%%%%%%%%%%%%%%%%%%%%%%%
        \begin{subfigure}[!t]{0.4\textwidth}
                \centering
                \includegraphics[width=\textwidth]{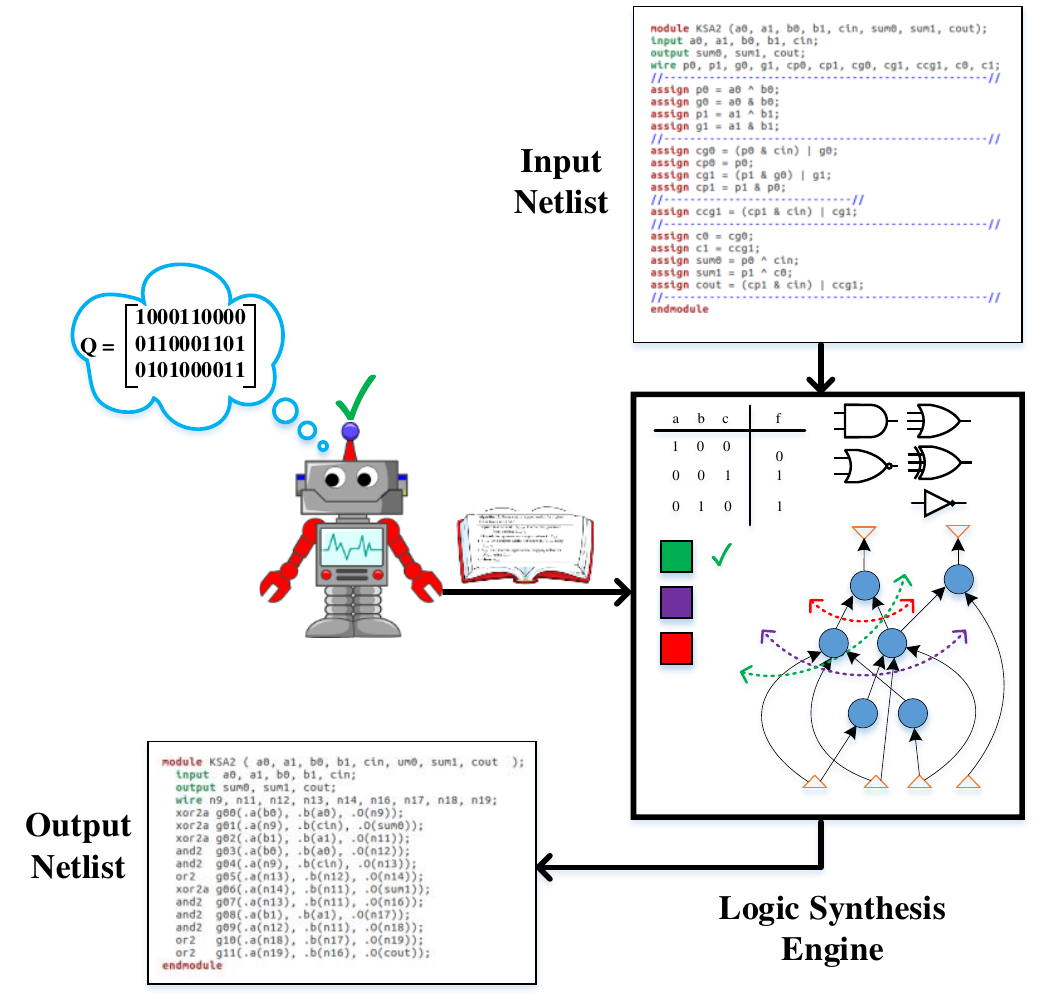}
                \caption{}
                \label{flow_b}
        \end{subfigure}
        \caption{Our Q-learning based ALS (Q-ALS) framework. (a) The Q-agent of Q-ALS is trained using the provided training networks and given the maximum error rate at primary outputs. (b) At the test time, this Q-agent helps synthesis tool to select the best approximate matches for implementing different nodes.}
        \label{overallapp}
\end{figure*}
%%%%%%%%%%%%%%%%%%%%%%%%%%%%%%%%%%%%%%%%%%%%%%%%%%%%%%%%%%%%%%%%%%%%%%%%%%%%%%%%%%%%%%%%
\section{Background}
\label{back:sec}
\subsection{Q-Learning}
\label{Q_Learning_sub_sec}
Q-learning is an RL algorithm that involves an agent, a set of states, and a set of actions. In Q-learning, at each time $t$ and state $s_t$, an action $a_t$ is taken, a reward $r_t$ is observed, and the agent enters a new state $s_{t+1}$. Q-learning algorithm can be modeled as a function that assigns a real number to each pairs of state-actions: $Q:S \times A \rightarrow \Bbb R$, and it can be expressed by a $Q$-matrix. 
The $Q$-matrix is updated using the following equation:
%%%%%%%%%%%%%%%%%%%%%%%%%%%%%%%%%%%%%%%%%%%%%%%%%%%%%%%%%%%%%%%%%%%
{\small
\begin{multline}
\label{Q_formula}
Q(s_t,a_t) \leftarrow (1 - \alpha) \times Q(s_{t-1},a_{t-1}) +  \\
\alpha \times \biggl(  r_t + \gamma \times \underset{a}{\operatorname{max}} ~Q(s_{t+1},a)   \biggr)
\end{multline}
}
%%%%%%%%%%%%%%%%%%%%%%%%%%%%%%%%%%%%%%%%%%%%%%%%%%%%%%%%%%%%%%%%%%%%%
where hyper-parameters $\alpha$ and $\gamma$ are \textit{learning rate} and \textit{discount factor}, respectively.  $\underset{a}{\operatorname{max}} ~Q(s_{t+1},a)$ is an estimation for optimal future reward, and  $r_t + \gamma \times \underset{a}{\operatorname{max}} ~Q(s_{t+1},a)$ is the learned value. 
Before the learning process starts, the $Q$-matrix is initialized into random values.
%%%%%%%%%%%%%%%%%%%%%%%%%%%%%%%%%%%%%%%%%%%%%%%%%%%%%%%%%%%%%%%%%%%%%%%%%%%%%%%%%%%%%%% 
\subsection{Cut-Based Technology Mapping}
\label{cut_based_subsec}
In state-of-the-art technology mappers \cite{pasandi2019pbmap, synthesis2011abc, pasandi2018sfqmap, mishchenko2005technology}, $k$-$feasible$ cuts  \cite{cong1994flowmap} are computed for each node in the given Boolean network, and subsequently, truth table of each cut is calculated. A truth table expresses the function of a cut based on its inputs. After computing truth tables, in a topological ordering traversal starting from level 1 nodes, the best cut for each node and the best supergate \cite{mishchenko2005technology} implementing this cut is computed. A supergate is a small combinational single output network built from original gates in the given library. After visiting all nodes, the best mapping solution (best cover) for the whole network is constructed by traversing the network back and using the computed best cuts and best implementation of those cuts.
%%%%%%%%%%%%%%%%%%%%%%%%%%%%%%%%%%%%%%%%%%%%%%%%%%%%%%%%%%%%%%%%%%%%%%%%%%%%%%%%%%%%%%%%%
%%%%%%%%%%%%%%%%%%%%%%%%%%%%%%%%%%%%%%%%%%%%%%%%%%%%%%%%%%%%%%%%%%%%%%%%%%%%%%%%%%%%%%%%%%%%%%%%%%%%%%%%%%%%%%%
\begin{figure}[t]
\centering
\includegraphics[width=0.5\textwidth]{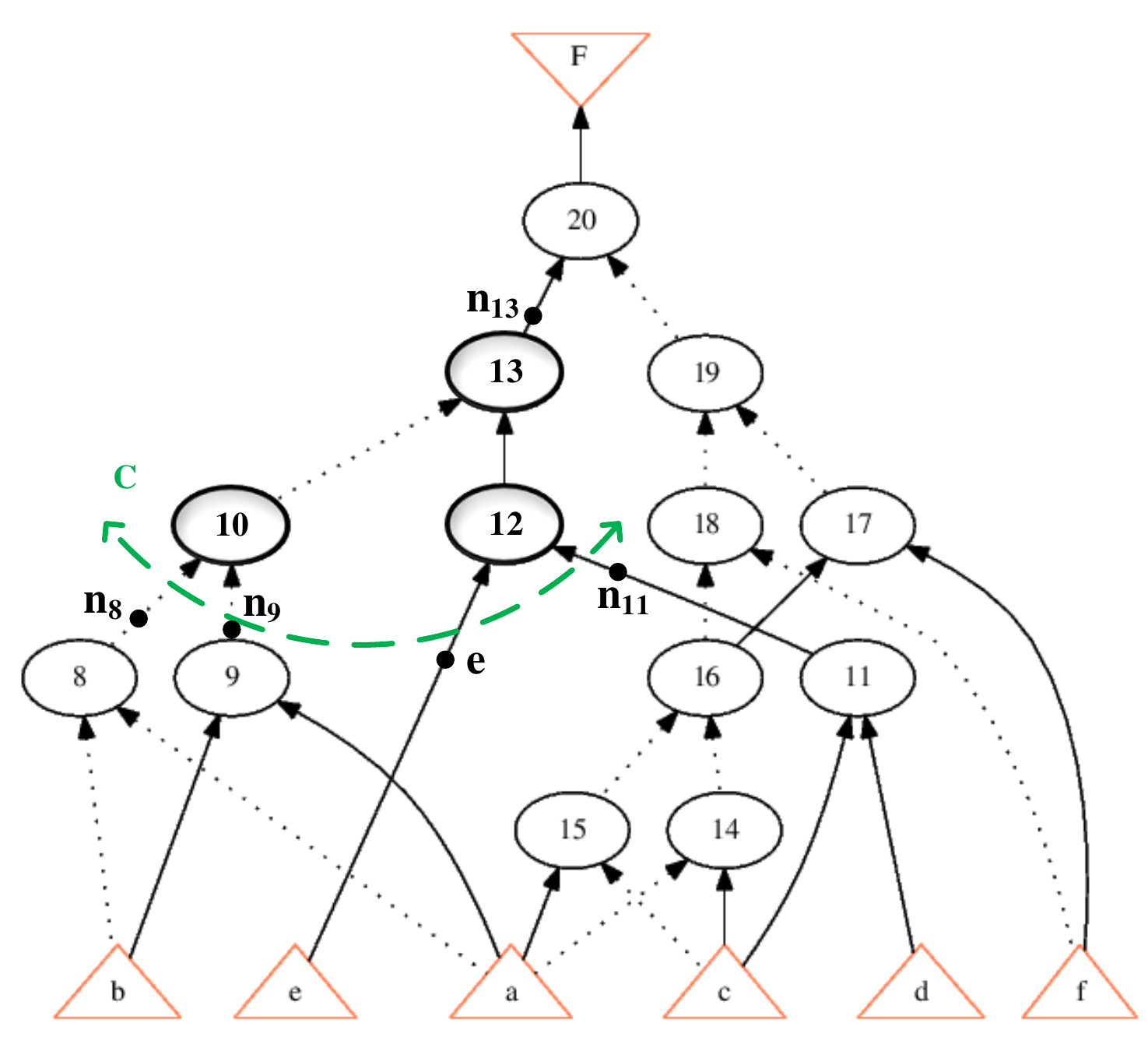}
\caption{An example network to demonstrate our method for finding approximate matches in a cut-based technology mapping approach. Function of node $n_{13}$ based on inputs of the shown cut is: $n_{13}$=$e \cdot n_{11} \cdot $ ($n_8$+$n_9$).}     
\label{Example_Network_fig}
\end{figure}
%%%%%%%%%%%%%%%%%%%%%%%%%%%%%%%%%%%%%%%%%%%%%%%%%%%%%%%%%%%%%%%%%%%%%%%%%%%%%%%%%%%%%%%%%%%%%%%%%%%%%%%%%%%%%%%
%%%%%%%%%%%%%%%%%%%%%%%%%%%%%%%%%%%%%%%%%%%%%%%%%%%%%%%%%%%%%%%%%%%%%%%%%%%%%%%%%%%%%%%%%%%%%%%%%%%%%%%%%%%%%%%%%%%%%%%
\begin{figure}[t]
        \centering
        \begin{subfigure}[!t]{0.24\textwidth}
                \centering
                \includegraphics[width=\textwidth]{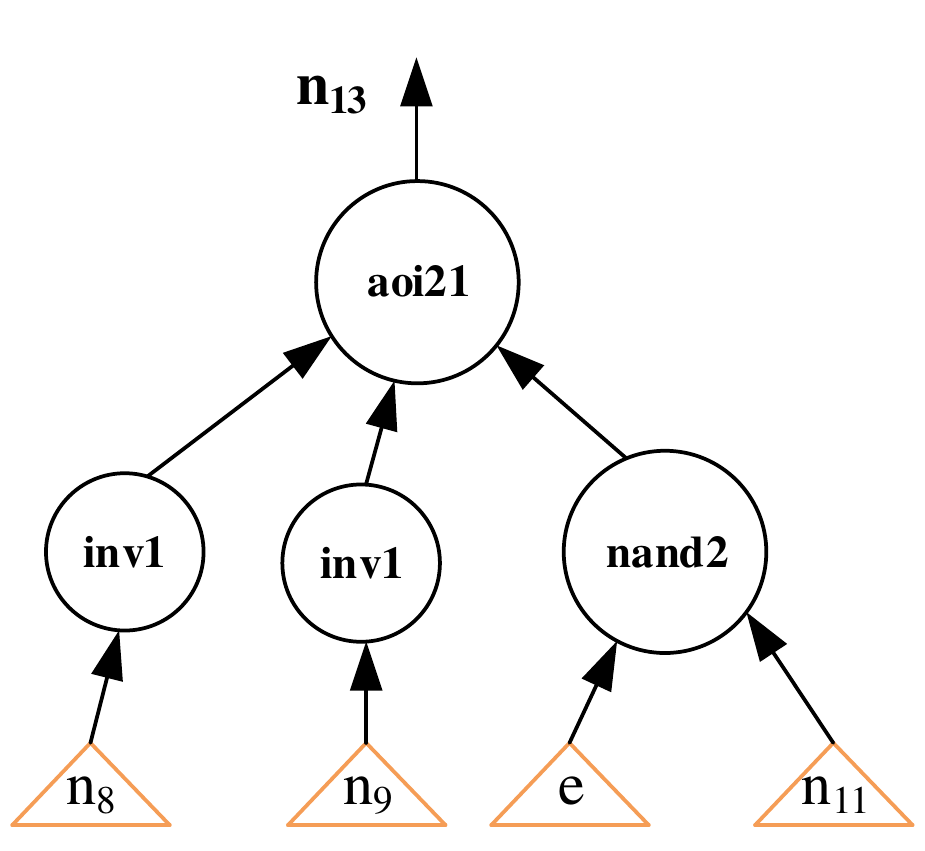}
                \caption{}
                \label{Implementation_1}
        \end{subfigure}
        %%%%%%%%%%%%%%%%%%%%%%%%%%
        \begin{subfigure}[!t]{0.24\textwidth}
                \centering
                \includegraphics[width=\textwidth]{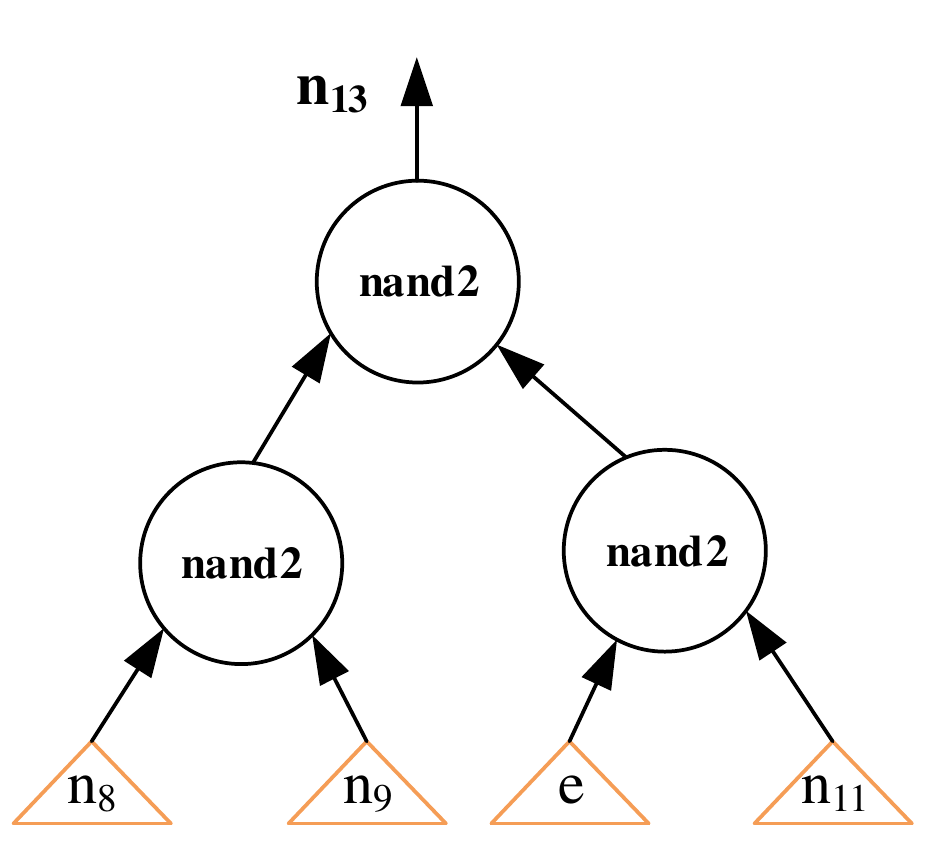}
                \caption{}
                \label{Implementation_2}
        \end{subfigure}
        \caption{Two implementations for function of cut $C$ the shown in Fig. \ref{Example_Network_fig}. (a) an exact implementation with area of 7.00 units and delay of 2.60 units (b) an approximate implementation with area of 6.00 units, delay of 2.00 units, and error rate of 37.5\%.}
        \label{Implementation_1_2}
\end{figure}
%%%%%%%%%%%%%%%%%%%%%%%%%%%%%%%%%%%%%%%%%%%%%%%%%%%%%%%%%%%%%%%%%%%%%%%%%%%%%%%%%%%%%%%%
\section{Our Proposed Framework for ALS: Q-ALS}
\label{prop:sec}
Fig. \ref{overallapp} illustrates Q-ALS, our proposed framework for ALS. In our Q-ALS, the cut-based technology mapping approach and the Q-learning algorithm are used to find the best approximate mapping solution for a given Boolean network. Given a maximum error rate at outputs of the given network, a Q-agent is trained using many training networks to learn the maximum error rate that each node can tolerate in order to maximize the delay and area savings. Hamming distance between truth tables of exact and approximate implementations is used as a metric for error rate, and to estimate the propagated error rate to outputs of the network resulted from approximation in a node, a Boolean difference calculus is used, which will be explained later in more details.   

For a given training network $N$, all nodes are visited in a topological ordering traversal and the exact and approximate mapping solutions (matches) to implement the best cut of each node are computed. Approximate matches are computed in two ways: (i) by simply dropping some gates from the exact match (ii) by examining other supergates in the supergate library that can implement function of the cut subject to a given maximum error rate. These approximate supergates are more cost efficient than the exact ones in terms of a desired metric such as delay, area, or power consumption. Similarly to the work in \cite{mishchenko2007combinational}, Q-ALS utilizes a prioritized cost function with delay having the highest priority while area is used as a tie breaker. 

Fig. \ref{Example_Network_fig} shows an example network to demonstrate our method for finding approximate matches in a cut-based technology mapping approach. Fig. \ref{Implementation_1} shows an exact implementation for the function of cut $C$ shown in Fig. \ref{Example_Network_fig}. $C$ is a cut among 4-feasible cuts of node $n_{13}$. The function of this node based on inputs of the shown cut is $n_{13}$=$e \cdot n_{11} \cdot $ ($n_8$+$n_9$). An approximate implementation with Hamming distance of six between the exact and approximate truth tables is shown in Fig. \ref{Implementation_2}. By dropping any of the shown inverters from the shown exact implementation, another approximate implementation with the same delay, one unit less area, and Hamming distance of two compared to the exact expression can be found. 
%%%%%%%%%%%%%%%%%%%%%%%%%%%%%%%%%%%%%%%%%%%%%%%%%%%%%%%%%%%%%%%%%%%%%%%%%%%%%%%%%%%%%%%%
\subsection{Action Space}
\label{action_sub_sec}
In Q-ALS, states and actions are defined as follows: nodes are considered as states, therefore, there are as many states as the total node count in And-Inverter Graph (AIG) \cite{mishchenko2004fraigs} representation of the given network. Given the current state as node $n$, the set of actions that can be taken are selecting a match for $n$ among different exact and approximate solutions for implementing k-feasible cuts of this node. A set of actions which do not generate a valid mapping solution for network $N$ are not desirable. For example, a set of actions is considered invalid, if those actions result in generating a mapping solution for the network $N$ such that the error rate at a primary output of this network exceeds the given maximum error rate.

To estimate the maximum error rate at primary outputs injected by approximate implementation of a node, we used the probabilistic error propagation approach presented in \cite{mohyuddin2011probabilistic}. This approach is based on Boolean difference calculus; it takes as input the Boolean function of a gate, error probabilities at its inputs, and the error probability of the gate itself, and produces the error probability at the output of this gate. We store propagated maximum error rate resulted from approximation in fanin cone of a node into data structure of this node. Next, for each choice of approximate implementation of a node and by using the said error propagation approach, we estimate the error rate at primary outputs of the network.

The Q-agent in Q-ALS learns the maximum error rate a single node can tolerate such that the saving in delay and then area is maximized. If the error rate on a single node is more than this value, it is estimated that it will violate the requirement of maximum error rate at primary outputs. Since the error metric that is employed in Q-ALS is Hamming distance, therefore, the Q-agent in Q-ALS basically learns the Maximum Hamming Distance (MHD) between exact and approximate truth tables of each node in the network.
%%%%%%%%%%%%%%%%%%%%%%%%%%%%%%%%%%%%%%%%%%%%%%%%%%%%%%%%%%%%%%%%%%%%%%%%%%%%%%%%%%%%%%%%%%%%%%%%%%%%%%%%%%%%%%%
\begin{figure}[b]
\centering
\includegraphics[width=0.45\textwidth]{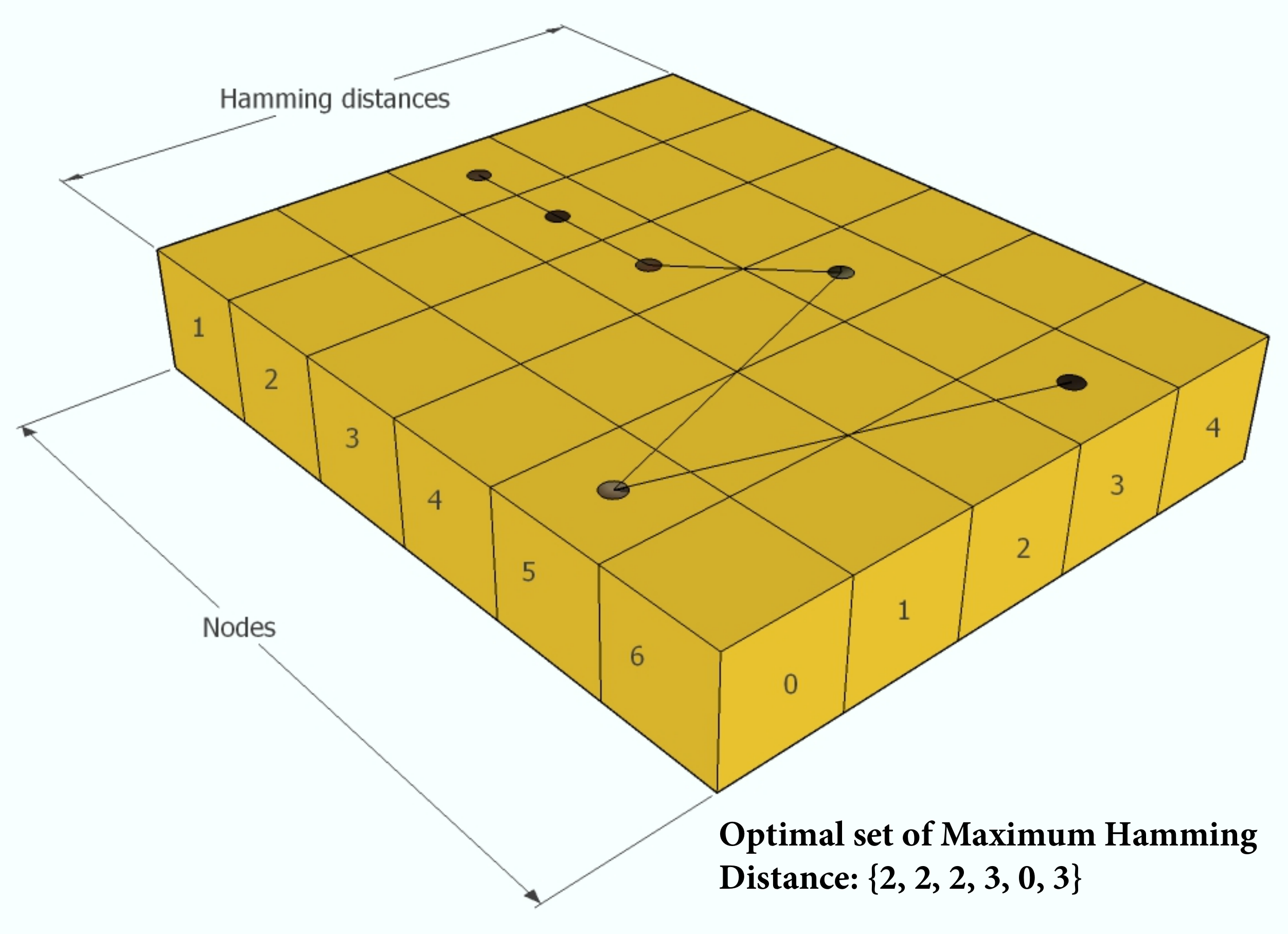}
\caption{A Q-matrix describing optimal set of Maximum Hamming Distance (MHD) for a network with six nodes. This network corresponds to implementing the following function: $F = a \oplus b \oplus c$. Note that $\oplus$ is an XOR operation.}     
\label{hamming_fig}
\end{figure}
%%%%%%%%%%%%%%%%%%%%%%%%%%%%%%%%%%%%%%%%%%%%%%%%%%%%%%%%%%%%%% 
\subsection{Reward Function}
\label{Reward_sub_sec}
A reward is assigned to each action based on the saving it offers for delay and area compared to the exact mapping solution. If an approximate match does not improve on the best delay or area of exact implementation, a reward of 0 is assigned to it. On the other hand, if it increases delay and/or area, a negative reward is assigned to it. The goal is to generate a \textit{valid} approximate mapping solution for $N$ which maximizes the total reward. 

Fig. \ref{hamming_fig} illustrates the Q-matrix for a network with six nodes and up to four valid exact and approximate mapping solutions for a single node. This network is for implementing the following function: $F = a \oplus b \oplus c$. One dimension of the corresponding box in this figure represents the nodes of the network, and the other dimension shows the MHD per node. Using the trained Q-agent, the following set of best MHDs are obtained for this network: $\{2,2,2,3,0,3\}$. This shows that for example the MHD between truth tables of exact and approximate solutions for node 1 is two. Therefore, a mapping solution with Hamming distance of zero, one, or two can be selected for implementing this node. MHD of zero corresponds to the exact mapping solution.
%%%%%%%%%%%%%%%%%%%%%%%%%%%%%%%%%%%%%%%%%%%%%%%%%%%%%%%%%%%%%%
\subsection{Training}
%%%%%%%%%%%%%%%%%%%%%%%%%%%%%%%%%%%%%%%%%%%%%%%%%%%%%%%%%%%%%%%%%%%%%%%%%%%%%%%%%%%%%%%%%%%%%%%%%%%%%
\begin{algorithm} [t]
{\small
\caption{Q-ALS model training using Q-learning}\label{algo:training}
\DontPrintSemicolon % Some LaTeX compilers require you to use \dontprintsemicolon instead
\KwIn{Set of training networks: $N_{train}=\{N_1,N_2,...,N_n\}$, Maximum error rate: $ER_{max}$ } 
\KwOut{Coefficients to get Maximum Hamming Distance (MHD) for a test network.}
{Initialize: \\  
\tab Number of Episodes: $N_E$ \\
\tab Learning rate: $\alpha$ \\
\tab Discount factor: $\gamma$ \\
\tab $Q$-matrix \\}
\For{each $N_i$ in $N_{train}$}{
Get number of nodes in $N_i$ \\
Assign random values for MHD of each node in $N_i$.\\
\For{each epoch in $N_E$}{
$N_{new}$ = Map $N_i$ using the assigned values for MHD. \\
\If{mapping is successful}{
$ER_{tmp}$=Find the maximum error rate at the outputs of $N_{new}$. \\
\If{$ER_{tmp}$ $<$ $ER_{max}$ and $Cost_{N_{new}}$ $<$ $Cost_{N_i}$}{
Update $Q$ matrix with positive reward. \\
Update $Cost_{N_i}$.}
\Else{Update $Q$-matrix with negative reward}
}
\Else{Update $Q$-matrix with negative reward}
}
}
Coefficients = Perform non-linear regression on $Q$-matrix \\
\Return{Coefficients}\;
}
\end{algorithm}
%%%%%%%%%%%%%%%%%%%%%%%%%%%%%%%%%%%%%%%%%%%%%%%%%%%%%%%%%%%%%%%%%%%%%%%%%%%%%%%%%%%%%%%%%%%%%%%%%%%%%%%%%%%%%
The training process starts with assigning a random integer number between 0 and 32 to MHD of each node. The maximum value of 32 comes from the fact that up to 5-cuts are computed for each node (32=$2^5$). These values are used to start the technology mapping process. If the generated mapping solution based on these MHD values is not valid, then the corresponding entry in the Q-matrix will be reduced. On the other hand, if the assigned MHDs for each node generate a network whose error rate is below the given maximum error rate, then the corresponding positions in the Q-matrix will be increased as long as it does not exceed the maximum value of 32. In following iterations, only a valid set of values which improves on the previous best delay and area values will update the Q-matrix. This process will continue for $N_E$ times for a particular network and will be repeated for other training networks.

The training process is shown in Algorithm \ref{algo:training}. Inputs of this algorithm are training networks and the maximum tolerable error rate at the outputs, $ER_{max}$. Values for hyper-parameters including number of episodes $N_E$, the learning rate $\alpha$, and the discount factor $\gamma$ are initialized inside this algorithm. Number of episodes is the number of times the training process is executed for a particular network. The Q-matrix generated during the training process will be a 2D matrix where $X$-axis denotes the nodes in the network and $Y$-axis denotes the number of options of Hamming distance for the corresponding node. For example, for a network with six nodes, the Q-matrix will have six entries in the $X$-axis, and a default of 32 entries for the $Y$-axis. 

At the end of the training, we will have a Q-matrix with a certain number of entries in the $X$-axis, which is equal to the maximum node count among training networks. Now, suppose that the node count in a test network is more than this value, then we cannot use this Q-matrix to test this network. To solve this issue, a non-linear regression is performed to fit a polynomial curve to the data in Q-matrix. Using this fitted curve, MHD values for nodes of any valid test network can be calculated.
%%%%%%%%%%%%%%%%%%%%%%%%%%%%%%%%%%%%%%%%%%%%%%%%%%%%%%%%%%%%%%%%%%%%%%%
\subsection{Testing}
The conclusion of training process returns coefficients of a polynomial function which will be used to predict the MHD values of a node in the given test network. These predicted MHDs will be used in the logic synthesis engine of Q-ALS to find the best approximate mapping solutions for individual nodes of the test network. The testing pseudo-code is shown in Algorithm \ref{algo:testing}.

%%%%%%%%%%%%%%%%%%%%%%%%%%%%%%%%%%%%%%%%%%%%%%%%%%%%%%%%%%%%%%%%%%%%%%%%%%%%%%%%%%%%%%%%%%%%%%%%%%%%%%%%%%%%%%%
\begin{algorithm} [t]
%{\small
\caption{Generating a mapped netlist for a given test network in Q-ALS}\label{algo:testing}
\DontPrintSemicolon % Some LaTeX compilers require you to use \dontprintsemicolon instead
\KwIn{Test network: $N_{Test}$, Coefficients generated from training: $L_{coeff}$  \\
 }
\KwOut{An approximate mapped network: $N_{opt}$}
$L_{HD}$ $<$= Compute MHDs for nodes in $N_{Test}$ using $L_{coeff}$. \\
$N_{opt}$ $<$= Generate approximate mapping netlist for $N_{Test}$ using $L_{HD}$.

\Return{$N_{opt}$}\;
%}
\end{algorithm}
%%%%%%%%%%%%%%%%%%%%%%%%%%%%%%%%%%%%%%%%%%%%%%%%%%%%%%%%%%%%%%%%%%%%%%%%%%%%%%%%%%%%%%%%%%%%%%%%%%%%%%%%%%%%%%%
\section{Experimental Results}
\label{exp:sec}
%%%%%%%%%%%%%%%%%%%%%%%%%%%%%%%%%%%%%%%%%%%%%%%%%%%%%%%%%%%%%%%%%%%%%%%%%%%%%%%%%%%%%%%%%%%%%%%%%%%%%%%%%%%%%%
\begin{table*}[ht]
\centering
\scriptsize
\caption{Experimental results for MCNC benchmark suite.}
\label{my-label}
\begin{tabular}{@{}cccccccc@{}}
\toprule
\textbf{Circuit} & \textbf{\begin{tabular}[c]{@{}c@{}}Node Count\end{tabular}} & \textbf{Exact Area} & \textbf{\begin{tabular}[c]{@{}c@{}}Exact Delay\end{tabular}} & \textbf{\begin{tabular}[c]{@{}c@{}}Approx. Area\end{tabular}} & \textbf{\begin{tabular}[c]{@{}c@{}}Approx. Delay\end{tabular}} & \textbf{Area Ratio} & \textbf{\begin{tabular}[c]{@{}c@{}}Delay  Ratio\end{tabular}} \\ \midrule
rd53     & 45     & 107    & 6.5   & 54  & 5.9   & 0.50   & 0.91 \\ 
\midrule
rd73  & 115   & 291  & 10.5  & 169  & 9.7   & 0.58   & 0.92 \\ 
\midrule
rd84 & 198  & 419 & 10.4 & 299  & 9.5 & 0.71  & 0.91  \\ 
\midrule
9sym   & 165 & 418  & 11.5  & 360  & 11.7 & 0.86   & 1.02  \\ 
\midrule
parity  & 15 & 75 & 7.6 & 24 & 4.9  & 0.32  & 0.64  \\ 
\midrule
my\_adder  & 149   & 361  & 36.4   & 161  & 26 & 0.45  & 0.71 \\ 
\midrule
z4ml  & 38   & 89 & 5.2   & 49   & 5.2  & 0.55 & 1.00  \\
\midrule
pm1  & 39   & 87   & 4.9   & 84   & 5.1  & 0.97   & 1.04  \\ 
\midrule
c8   & 122    & 281   & 6.9  & 176   & 6.4  & 0.63  & 0.93   \\ 
\midrule
x4   & 339 & 760    & 7.5  & 760  & 7.9   & 1.00   & 1.05  \\
\midrule
count    & 123 & 273    & 13.8   & 171   & 12   & 0.63     & 0.87   \\ 
\midrule
pcler8  & 59   & 150  & 7.7  & 126  & 7.2    & 0.84     & 0.94 \\ 
\midrule
sct    & 68   & 151    & 5.5   & 137   & 5.7  & 0.91  & 1.04  \\ 
\midrule
apex7   & 187   & 409   & 12.3  & 368    & 10  & 0.90    & 0.81   \\ 
\bottomrule
\label{Table_1}
\end{tabular}
\end{table*}
%%%%%%%%%%%%%%%%%%%%%%%%%%%%%%%%%%%%%%%%%%%%%%%%%%%%%%%%%%%%%%%%%%%%%%%%%%%%%%%%%%%%

% \usepackage{booktabs}
\begin{table*}[ht]
\scriptsize
\centering
\caption{Experimental results for ITC 99 benchmark suite.}
\label{my-label}
\begin{tabular}{@{}ccccccccc@{}}
\toprule
\textbf{Circuit} & \textbf{Function}                                 & \textbf{\begin{tabular}[c]{@{}c@{}}Node Count\end{tabular}} & \textbf{Exact Area} & \textbf{\begin{tabular}[c]{@{}c@{}}Exact Delay\end{tabular}} & \textbf{\begin{tabular}[c]{@{}c@{}}Approx. Area\end{tabular}} & \textbf{\begin{tabular}[c]{@{}c@{}}Approx. Delay\end{tabular}} & \textbf{Area Ratio} & \textbf{\begin{tabular}[c]{@{}c@{}}Delay Ratio\end{tabular}} \\ \midrule
b01              & FSM comparing serial flows & 34  & 86                  & 4.7                                                              & 56                                                                             & 5.7                                                                            & 0.65                & 1.21                                                             \\ \midrule
b02              & FSM that recognizes BCS numbers                   & 19                                                                & 40                  & 5                                                                & 19                                                                             & 2.9                                                                            & 0.48                & 0.58                                                             \\ \midrule
b04              & Compute min and max                               & 392                                                               & 1018                & 18.6                                                             & 808                                                                            & 14.8                                                                           & 0.79                & 0.80                                                             \\ \midrule
b06              & Interrupt handler                                 & 34                                                                & 86                  & 4.1                                                              & 45                                                                             & 4.3                                                                            & 0.52                & 1.05                                                             \\ \midrule
b07              & Count points on a straight line                   & 274                                                               & 671                 & 18.4                                                             & 559                                                                            & 16.5                                                                           & 0.83                & 0.90                                                             \\ \midrule
b08              & Find inclusions in sequences of numbers           & 137                                                               & 308                 & 12.2                                                             & 259                                                                            & 10.1                                                                           & 0.84                & 0.83                                                             \\ \midrule
b09              & Serial to serial converter                        & 118                                                               & 323                 & 8.3                                                              & 233                                                                            & 8.6                                                                            & 0.72                & 1.04                                                             \\ \midrule
b10              & Voting system                                     & 142                                                               & 342                 & 9                                                                & 276                                                                            & 8.9                                                                            & 0.81                & 0.99                                                             \\ \midrule
b11              & Scramble string with variable cipher              & 441                                                               & 1119                & 20.1                                                             & 828                                                                            & 13.7                                                                           & 0.74                & 0.68                                                             \\ \midrule
b12              & 1-player game (guess a sequence)                  & 843                                                               & 1993                & 11.8                                                             & 1642                                                                           & 12.1                                                                           & 0.82                & 1.03                                                             \\ \midrule
b13              & Interface to meteo sensors                        & 223                                                               & 517                 & 8.7                                                              & 372                                                                            & 7.9                                                                            & 0.72                & 0.91                                                             \\ \midrule
b14              & Viper processor (subset)                          & 4233                                                              & 10688               & 44.9                                                             & 9361                                                                           & 40.4                                                                           & 0.88                & 0.90                                                             \\ \midrule
b17              & Three copies of 80386 processor (subset)          & 18446                                                             & 47667               & 69.5                                                             & 41419                                                                          & 60.5                                                                           & 0.87                & 0.87                                                             \\ \midrule
b18              & Two copies of b14 and two of b17                  & 57978                                                             & 145896              & 105.5                                                            & 136755                                                                         & 105.5                                                                          & 0.94                & 1.00                                                             \\ \midrule
b20              & A copy and a modified version of b14       & 8805                                                              & 21598               & 53.1                                                             & 17303                                                                          & 45.6                                                                           & 0.80                & 0.86                                                             \\ \midrule
b21              & Two copies of b14                                 & 9141                                                              & 22705               & 52.9                                                             & 19060                                                                          & 44.8                                                                           & 0.84                & 0.85                                                             \\ \midrule
b22              & A copy and two modified versions of b14    & 13561                                                             & 33260               & 53.3                                                             & 26573                                                                          & 46.5                                                                           & 0.80                & 0.87         \label{Table_2}                                                    \\ \bottomrule
\end{tabular}
\end{table*}
%%%%%%%%%%%%%%%%%%%%%%%%%%%%%%%%%%%%%%%%%%%%%%%%%%%%%%%%%%%%%%%%%%%%%%%%%%%%%%%%%%%%%%%%%%%%%%%%%%%%%%%%%%%%%%%%%%%%%%%%%%%%%%%%%%%%%%
We implemented Q-ALS as an extension to ABC \cite{synthesis2011abc}. We first trained our model on the ISCAS 89\cite{100747} and EPFL\cite{amaru2015epfl} benchmark circuits and computed entries of the Q-matrix. The maximum error rate at primary outputs used in the experimental results presented in this section is 5\%, which is the same as two other state-of-the-art ALS frameworks (i.e., SASIMI and Single/Multi-Selection) compared in this section. The generic standard cell library, \textit{mcnc.genlib}, consisting of 25 gates is used in technology mapping. Test circuits are chosen from different benchmark suites including ISCAS 85\cite{iscas}, MCNC\cite{yang1991logic}, and ITC 99\cite{corno2000rt}. The functionality of circuits widely varies from simple arithmetic circuits of EPFL benchmark suite to complex industry-level circuits of ITC 99 benchmark suite. All experiments were conducted on a virtual machine running Linux with 1GB RAM and a 2.4 GHz laptop as the host machine. Tables \ref{Table_1}$-$\ref{Table_3} contain list experimental results for different circuits. The complexity of circuits in terms of number of gates, exact area, and exact delay is also shown.

In Table \ref{Table_1}, the 3rd and 4th columns represent exact values of area and delay, respectively.  These values are obtained by ABC for benchmark circuits in the first column. The second column represents the number of gates in the corresponding benchmark circuit. Columns five and six represent the area and delay of approximated circuits generated by Q-ALS. Columns seven and eight list area and delay ratios. Area ratio is calculated by dividing the area of an approximate circuit by the area of the corresponding exact circuit. Similarly, the delay ratio is calculated. 

Table \ref{Table_1} shows the area and delay comparison of approximated circuits with their exact counterparts in the MCNC benchmark suite. The MCNC benchmark suite contains different types of circuits such as Finite State Machine (FSM) circuits, multi-level combinational circuits, multi-level sequential circuits, and two-level circuits. Experimenting on these circuits, we get up to 70\% area reduction and up to 36\% delay reduction using our Q-ALS. The average area and delay reduction as compared to the baseline values (exact solutions) are 30\% and 9\%, respectively.

%%%%%%%%%%%%%%%%%%%%%%%%%%%%%%%%%%%%%%%%%%%%%%%%%%%%%%%%%%%%%%%%%%%%%%%%%%%%%%%%%%%%%%%%%%%%%%%%%%%%%%%%%%%%%%%%%%%%%%%%%%%%%%%%%%%%%%%%%%%%%%%%%%
\begin{table*}[ht]
\centering
\scriptsize
\caption{Experimental results for ISCAS 85 benchmark suite.}
\label{my-label}
\begin{tabular}{@{}cccccccccc@{}}
\toprule
\textbf{}        & \textbf{}           & \multicolumn{2}{c}{\textbf{SASIMI \cite{venkataramani2013substitute}}}                     & \multicolumn{2}{c}{\textbf{Single-Selection \cite{wu2016efficient}}}  & \multicolumn{2}{c}{\textbf{Multi-Selection \cite{wu2016efficient}}}   & \multicolumn{2}{c}{\textbf{Q-ALS}}               \\ \midrule
\textbf{Circuit} & \textbf{Exact Area} & \textbf{Approx. Area} & \textbf{Area ratio} &\textbf{Approx. Area} & \textbf{Area ratio}& \textbf{Approx. Area} & \textbf{Area ratio} & \textbf{Approx. Area} & \textbf{Area ratio} \\ \midrule
c880    & 646  & 579   &  0.896  &    577   &  0.893   &  577   & 0.893   & 558  & 0.864   \\
 \midrule
c1908    & 846     &  516   &  0.610  &  503  & 0.595  &   506  & 0.598   & 520  &  0.615 \\ 
\midrule
c2670   & 1298      &  940  &  0.724   &  859  & 0.662   &   874  & 0.673  &  866 & 0.667  \\ 
\midrule
c3540     & 1916    &  1868 & 0.975  &   1851  & 0.966    &  1849  & 0.965  &  1812  & 0.945  \\ 
\midrule
c5315      & 3060   &  3002  & 0.981  &  2993  & 0.978  &  3002  & 0.981   & 1901   &  0.621  \\ 
\midrule
c7552      & 3952   &  3746  & 0.948   & 3715  & 0.940  &  3719  & 0.941  &  2451  &  0.620       \\ 
\midrule
alu4      & 2740    &  2444  & 0.892  &  2406  & 0.878  & 2381   &  0.869  & 1260  &  0.460      \\ 
\bottomrule
\label{Table_3}
\end{tabular}
\end{table*}
%%%%%%%%%%%%%%%%%%%%%%%%%%%%%%%%%%%%%%%%%%%%%%%%%%%%%%%%%%%%%%%%%%%%%%%%%%%%%%%%%%%%%%%%%%%%%%%%%%%%%%%%%

Apart from academic benchmark circuits, we also experimented on industry-level benchmarks from ITC 99 benchmark suite \cite{corno2000rt}. Table \ref{Table_2} shows experimental results for this benchmark suite together with a short description for functionality of each circuit. On average for 17 ITC 99 benchmark circuits, Q-ALS provides 23\% reduction in area and 10\% reduction in delay compared to the exact solution (baseline). This shows that our proposed methodology is quite effective in area and delay reduction for industrial-level circuits as well.

We experimented on benchmark circuits of ISCAS 85 benchmark suite to compare the results of our proposed framework (Q-ALS) with state-of-the-art approximate logic synthesis tool: SASIMI \cite{venkataramani2013substitute} and Single/Multi-Selection approaches \cite{wu2016efficient}. The reason behind choosing ISCAS 85 for this comparison is availability of experimental results in both of these papers for these benchmarks. Table \ref{Table_3} shows the area comparison between SASIMI, Single/Multi-Selection, and our Q-ALS for ISCAS 85 benchmark circuits. We observed that the average area reduction by SASIMI compared with the exact values is 13.9\% and for Single and Multi-Selection are 15.6\%, and 15.5\%, respectively. The average area reduction by Q-ALS is 31.6\% which clearly outperforms other counterparts.
%%%%%%%%%%%%%%%%%%%%%%%%%%%%%%%%%%%%%%%%%%%%%%%%%%%%%%%%%%%%%%%%%%%%%%%%%%%%%%%%%%%%%%%%%%%%%%%%%%%%%%%%
\begin{figure}[b]
\centering
\includegraphics[width=0.5\textwidth]{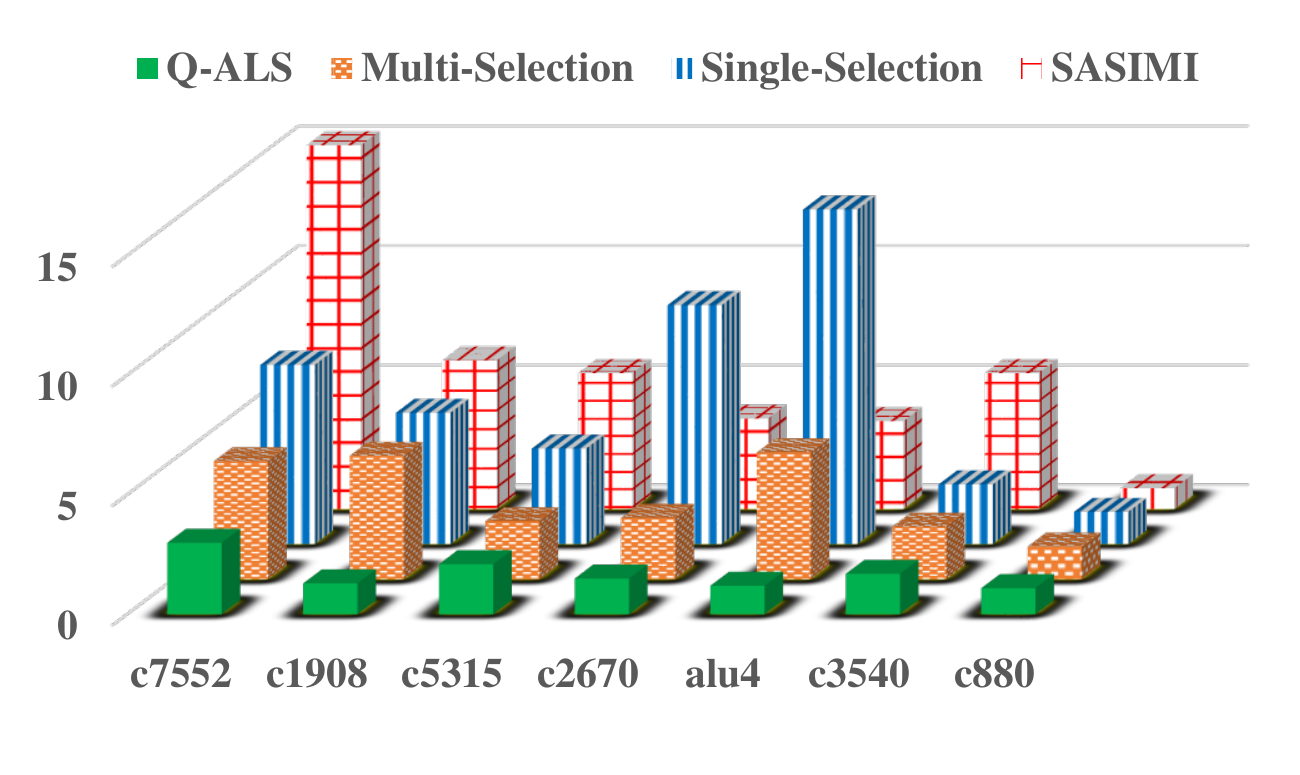}
\caption{Run-time (in second) comparison among SASIMI, Single/Multi-Selection, and our Q-ALS for approximate logic synthesis. For better exhibition purposes, data for Single-Selection and SASIMI are scaled down by 5 and 2 times, respectively.}
\label{run_time_fig}
\end{figure}        
%%%%%%%%%%%%%%%%%%%%%%%%%%%%%%%%%%%%%%%%%%%%%%%%%%%%%%%%%%%%%%%%%%%%%%%%%%%%%%%%%%%%%%%%%%%%%%%%%%%%%%%%

Fig. \ref{run_time_fig} shows the comparison of run-times of Q-ALS with SASIMI and Single/Multi-Selection. We observed that on average, run-time of Q-ALS as compared to SASIMI, Single-Selection, and Multi-Selection is 15.94$\times$, 8.46$\times$, and 2.21$\times$ less, respectively. This is because the inference time of the Q-learning algorithm is very low. The time required to train our model using Q-learning algorithm is around five hours. However, once the model is trained, it can generate the approximated mapping solutions for any given circuit in a small amount of time. In other words, it is trained once, but is used for generating approximate mapping solutions many times. For this reason, the training time is not considered in Fig. \ref{run_time_fig}. Please note that SASIMI and Single/Multi-Selection algorithms are implemented in SIS \cite{sentovich1992sis}, while Q-ALS is implemented in ABC, which is much faster than SIS. To remove this bias, we normalized run-time of SASIMI and Single/Multi Selection approaches by a factor obtained from data published in \cite{mishchenko2005technology,mishchenko2006dag}.

\section{Conclusion}
\label{conc:sec}
In this paper, we present Q-ALS, a reinforcement learning based approximate logic synthesis framework. Q-ALS benefits from strong capabilities of Q-learning algorithm to learn the maximum error rate each node in the AIG form of the given network can tolerate in order to achieve a maximum saving in delay and area while bounding to a predetermined error rate at the primary outputs. Thanks to this capability, Q-ALS is able to provide up to 70\% area reduction and 36\% delay reduction for academic benchmarks, and up to 52\% area reduction and 30\% delay reduction for industrial-level benchmarks. Furthermore, Q-ALS reduces run-time by an average of 15.94$\times$ and 2.21$\times$ over SASIMI and Multi-Selection, two academic state-of-the-art ALS tools. 
\ifCLASSOPTIONcaptionsoff
  \newpage
\fi
\bibliographystyle{IEEEtran}
\bibliography{IEEEabrv,Q-ALS}

\end{document}